\documentclass[twocolumn,prd]{revtex4}
\usepackage{epsfig}

\begin{document}

\title{Model-independent search for the Abelian $Z'$ boson in the Bhabha process}
\author{A.V. Gulov and V.V. Skalozub}
\affiliation{Dniepropetrovsk National University, Ukraine}
\begin{abstract}
Model-independent observables to pick out  the Abelian $Z'$ signal
in the Bhabha process are introduced at energies $\sqrt{s}\ge 200$
GeV. They measure separately the $Z'$-induced vector and
axial-vector four-fermion contact couplings. The analysis of the
LEP2 data constrains the value of the $Z'$-induced vector
four-fermion coupling at the $2\sigma$ confidence level that
corresponds to the Abelian $Z'$ boson with the mass of order 1
TeV.
\end{abstract}
\maketitle

\section{Introduction}

The LEP2 experiments have successfully confirmed the predictions
of the Standard model (SM). At the same time, they inspire
numerous estimations of possible new heavy particles beyond the
SM. In particular, searching for ``new physics'' has already
become an obligatory part of the reports on modern experiments in
high energy physics. To describe possible signals of new physics
beyond the SM the LEP2 collaborations apply both the
model-dependent and the model-independent analysis of experimental
data. The former approach means the comparison of experimental
data with the predictions of some specific models which extend the
SM at high energies. In this way a number of grand unified
theories, the supersymmetry models were discussed and their
parameters have been restricted. These model-dependent bounds are
adduced  in the reports \cite{EWWG}.

In the model-independent approach one fits some low-energy
parameters such as the four-fermion contact couplings. To establish a
model-independent fit one has to take into consideration all
possible coupling constants of a new heavy particle with the SM
particles and  introduce the observables which pick  only one of
these parameters. To find the expected signals the LEP
collaborations have applied a ``helicity model fit''. In this
analysis an effective Lagrangian describing contact interactions
of massless fermion states with one specific helicity (axial-axial
(AA) model, vector-vector (VV) model, etc.) was introduced and the
corresponding couplings have been restricted. This approach,
giving a possibility to detect the signals of new physics, does
not discern the specific state (its quantum numbers) responsible
for them. This is because the particle interactions are described by
a number of structures which contribute to different ``helicity
models''. Therefore a specific particle contributes to a number of
the ``models''.

In this regard, it seems resonable to develop an approach allowing
to pick out in a model-independent way the parameters of new heavy
particles with specified quantum numbers. In the previous papers
\cite{PRD00,ZprTHDM,YAF04} we established a model-independent
search for manifestations of a heavy Abelian $Z'$ boson beyond the
SM. The key point of this analysis was the fact that for any
theory beyond the SM which is renormalizable one (but unspecified
in other respects) some relations between the unknown low energy
$Z'$ couplings with fermions hold \cite{PRD00}. In particular,
they require an universal value of the $Z'$ couplings to the
axial-vector fermion currents. Taking into account these
model-independent relations as well as the kinematics features of
the processes $e^+e^-\to\mu^+\mu^-,\tau^+\tau^-$, we introduced
the unique sign-definite observable to select the signal of the
virtual $Z'$ state \cite{PRD00}. The value of the observable
measures the $Z'$-induced four-fermion coupling of the
axial-vector currents. The analysis of the LEP2 data set on
$e^+e^-\to\mu^+\mu^-,\tau^+\tau^-$ processes has  shown that the
mean value of the observable is in an accordance with the signal
although the accuracy is at the 1$\sigma$ confidence level (CL).
Clearly, this could not be regarded as the actual observation of
the particle.

As numerous estimates for various scattering processes showed the
LEP2 data set is not too large to detect the signals of new
physics at more then the $1\sigma$ CL (see, for instance, Refs.
\cite{EWWG,bourilkov}). Such an accuracy is insufficient for a
real discovery. However, one may believe that the signals would
appear more evidently if the statistics increases. One of the ways
to ensure that within the existing data set is to follow the third
of the described approaches and investigate other processes where
the virtual states of the chosen new heavy particles contribute
and can be identified. If again the observables to single out them
are introduced and the data set is treated accordingly, one
obtains an independent information about the states. Then,
dependently on the results, one is able to conclude about the
existence of the states and their characteristics by accounting
for both of experiments that increases statistics.

These speculations served as a motivation for investigations
carried out in the present paper in order to search for the
Abelian $Z'$ gauge boson in the Bhabha processes
$e^+e^-\rightarrow e^+e^-(\gamma)$. We introduce new
model-independent observables sensitive to the vector, $v_f$, and
axial-vector, $a_f$, $Z'$ couplings to the electron (positron)
currents. They can be constructed from differential cross-sections
at the LEP2 energies as well as at the energies of future
electron-positron colliders ($\geq 500$ GeV). As an application,
we analyse the existing in the literature LEP2 data on the Bhabha
process and derive limits on fitted parameters. The comparison
with the results obtained already for the $e^+e^-\rightarrow
\mu^+\mu^-,\tau^+\tau^-$ processes is done. We also confront our
results with other analysises for Bhabha process. The paper is
organized as follows. In sect. II we introduce the effective
Lagrangian describing the $Z'$-boson interaction with fermions at
low energies as well as other necessary information on the
model-independent search for this state at low energies. In sect.
III we investigate the differential cross-sections with taking
into consideration the relations between $a_f$ and $v_f$
parameters peculiar for the Abelian $Z'$ that gives a possibility
to identify this virtual state. In sects. IV, V the observables to
pick out $v^2_f$ and $a^2_f$ are introduced. The last two sections
are devoted to the analysis of the LEP2 data and discussion.

\section{The Abelian $Z'$ couplings to the SM particles}

The $Z'$-boson can be introduced in a phenomenological way by
specifying its effective low-energy couplings to the known SM
particles \cite{leike}. Considering the $Z'$ effects at energies
much below the $Z'$ mass, it is enough to parametrize the
tree-level $Z'$ interactions of renormalizable types, only. Such a
possibility is provided by the decoupling theorem
\cite{decoupling} from which, in particular, it follows that the
non-renormalizable $Z'$ interactions are produced by loops at
higher energies and suppressed by powers of the inverse $Z'$ mass.
The SM gauge group $SU(2)_L\times U(1)_Y$ is usually considered as
a subgroup of an underlying theory gauge group, so the
vector-boson interactions of types $Z'W^+W^-$, $Z'ZZ$, ... are
absent at a tree level \cite{arzt}. Thus, to investigate the $Z'$
effects in the leptonic processes at the energies of LEP
experiments ($\sqrt{s}\sim 200$ GeV) or future electron-positron
colliders ($\sqrt{s}\ge 500$ GeV) we use the Lagrangian:
\begin{eqnarray}\label{1}
 {\cal L}&=& \left|\left( D^{{\rm ew,} \phi}_\mu -
  \frac{i\tilde{g}}{2}\tilde{Y}(\phi)\tilde{B}_\mu
  \right)\phi\right|^2 + \nonumber\\&&
 i\sum\limits_{f=f_L,f_R}\bar{f}{\gamma^\mu}
  \left(
  D^{{\rm ew,} f}_\mu -
  \frac{i\tilde{g}}{2}\tilde{Y}(f)\tilde{B}_\mu
  \right)f,
\end{eqnarray}
where $\phi$ is the SM scalar doublet, $\tilde{B}_\mu$ denotes the
$Z'$ field before the spontaneous breaking of the electroweak
symmetry, and summation over the all SM left-handed fermion
doublets, $f_L =\{(f_u)_L, (f_d)_L\}$, and the right-handed
singlets, $f_R = (f_u)_R, (f_d)_R$, is assumed. The notation
$\tilde{g}$ stands for the charge corresponding to the $Z'$ gauge
group, $D^{{\rm ew,}\phi}_\mu$ and $D^{{\rm ew,}f}_\mu$ are the
electroweak covariant derivatives. Diagonal $2\times 2$ matrices
$\tilde{Y}(\phi)={\rm
diag}(\tilde{Y}_{\phi,1},\tilde{Y}_{\phi,2})$,
$\tilde{Y}(f_L)={\rm diag}(\tilde{Y}_{L,f_u},\tilde{Y}_{L,f_d})$
and numbers $\tilde{Y}(f_R)=\tilde{Y}_{R,f}$ mean the unknown $Z'$
generators characterizing the model beyond the SM.

The Lagrangian (\ref{1}) generally leads to the $Z$--$Z'$ mixing
of order $m^2_Z/m^2_{Z'}$ which is proportional to
$\tilde{Y}_{\phi,2}$ and originated from the diagonalization of
the neutral vector boson mass matrix. This mixing contributes to
the scattering amplitudes and plays an important role at the LEP2
energies \cite{NPB}.

The low energy $Z'$ couplings to a fermion $f$ are parameterized
by two numbers $\tilde{Y}_{L,f}$ and $\tilde{Y}_{R,f}$.
Alternatively, the couplings to the axial-vector and vector
fermion currents,
$a^l_{Z'}\equiv(\tilde{Y}_{R,l}-\tilde{Y}_{L,l})/2$ and
$v^l_{Z'}\equiv(\tilde{Y}_{L,l}+\tilde{Y}_{R,l})/2$, are often
used. Their specific values are determined by the unknown model
beyond the SM. Assuming an arbitrary underlying theory one usually
supposes the parameters $a_f$ and $v_f$ are to be independent
numbers. However, as it was proved in Ref. \cite{ZprTHDM}, for any
renormalizable theory beyond the SM these parameters content some
relations which follow from the renorlalization group equations
and the decoupling theorem.  In case of the $Z'$ boson this is
reflected in  correlations between $a_f$ and $v_f$. These
correlations are model-independent in sense that they do not
depend on an particular underlying model. The detailed discussion
of these issues and the derivation of the RG relations are
presented in Ref. \cite{ZprTHDM}. Therein, in particular, it is
shown that the low energy Abelian $Z'$ couplings are constrained
by the relations
\begin{eqnarray}\label{2}
& v_f-a_f=v_{f^\star}-a_{f^\star},\quad
\tilde{Y}_{\phi,1}=\tilde{Y}_{\phi,2}\equiv\tilde{Y}_\phi,&
\nonumber\\ & a_f=T_{3,f}\tilde{Y}_\phi,&
\end{eqnarray}
where $T^3_f$ is the third component of the fermion weak isospin,
and $f^\star$ means the isopartner of $f$ (namely,
$l^\star=\nu_l,\nu^\star_l=l,\ldots$).

The correlations (\ref{2}) result in important properties of the
Abelian $Z'$ couplings. They ensure, in particular, the invariance
of the Yukawa terms with respect to the effective low-energy
$\tilde{U}(1)$ subgroup corresponding to the Abelian $Z'$ boson.
As it follows from the relations, the couplings of the Abelian
$Z'$ to the axial-vector fermion currents have the universal
absolute value proportional to the $Z'$ coupling to the scalar
doublet. So, we will use the short notation
$a=a_l=-\tilde{Y}_\phi/2$.

The relations (\ref{2}) allow to reduce the number of independent
parameters of new physics. The leading-order differential
cross-section of the Bhabha process depends on five different
combinations of $Z'$ couplings: $a_e^2$, $v_e^2$, $a_e v_e$, $a_e
\tilde{Y}_\phi$, and $v_e \tilde{Y}_\phi$. Taking into account
these correlations reduce them to three actual parameters $a_e^2$,
$v_e^2$, and $a_e v_e$. As it will be shown, the factors at
$a_e^2$ and $v_e^2$ are dominant. Therefore, with a high accuracy
the differential cross-section is a two-parametric function.

In Refs. \cite{PRD00,YAF04} we showed for the processes
$e^+e^-\to\mu^+\mu^-,\tau^+\tau^-$, that the many-parametric
differential cross section can be integrated over a special range
of the scattering angles in order to select only one of the
unknown parameters. It seems resonable to apply similar approach
in order to measure separately the couplings $a_e^2$ and $v_e^2$
by using the differential cross-section of the Bhabha process.
Such kind of observables can be fitted by means of treating of
modern experimental data that  provides the bounds on the low
energy vector and the axial-vector $Z'$ couplings to electron. Due
to the mentioned above universality of the axial-vector coupling
$a^2$ it is possible to compare the estimates derived in different
independent processes.

\section{Cross-section}

\begin{figure}
\centering
\includegraphics[bb= 0 0 600 400 ,width=70mm]{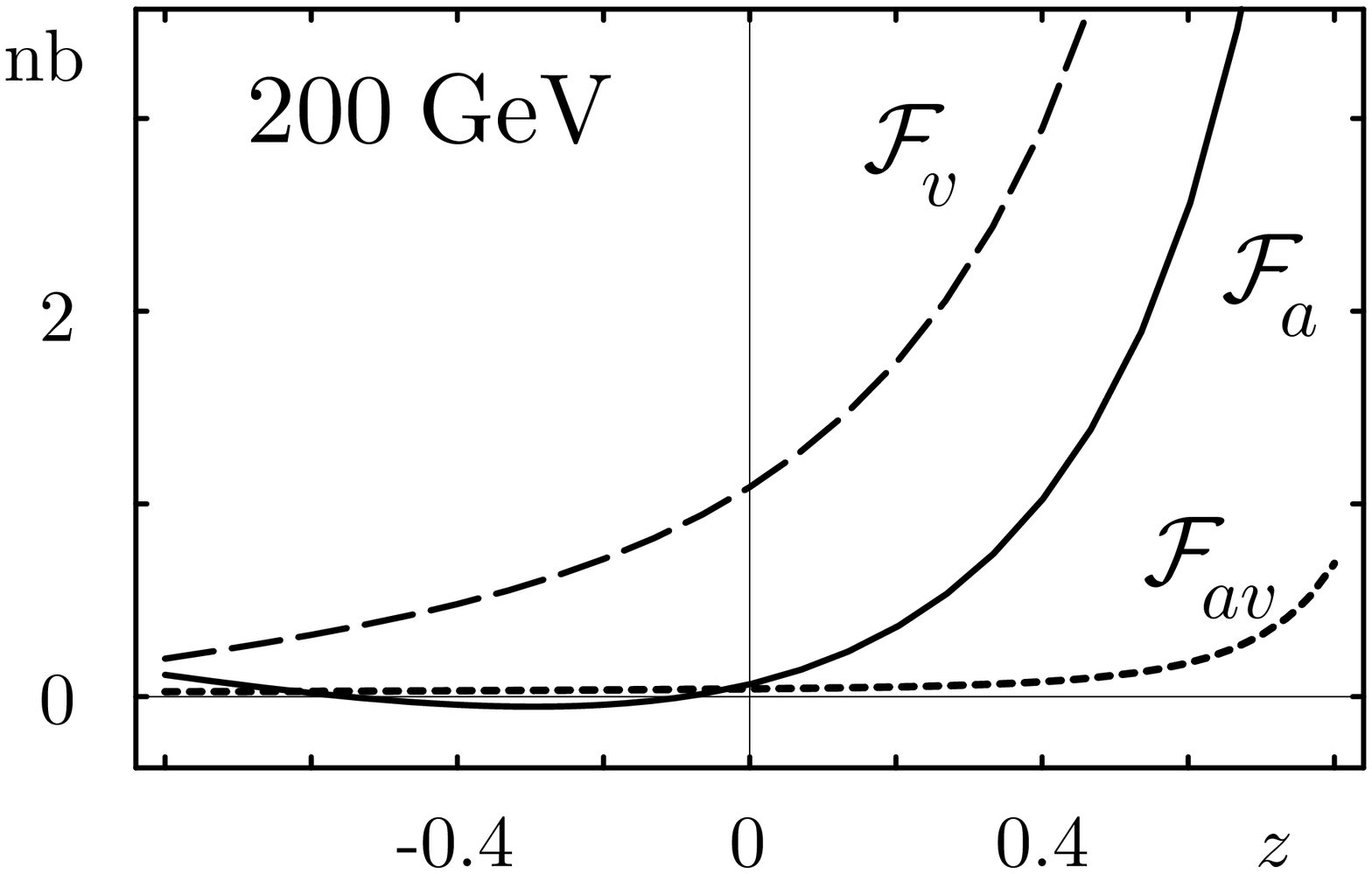}
\\
\includegraphics[bb= 0 0 600 400 ,width=70mm]{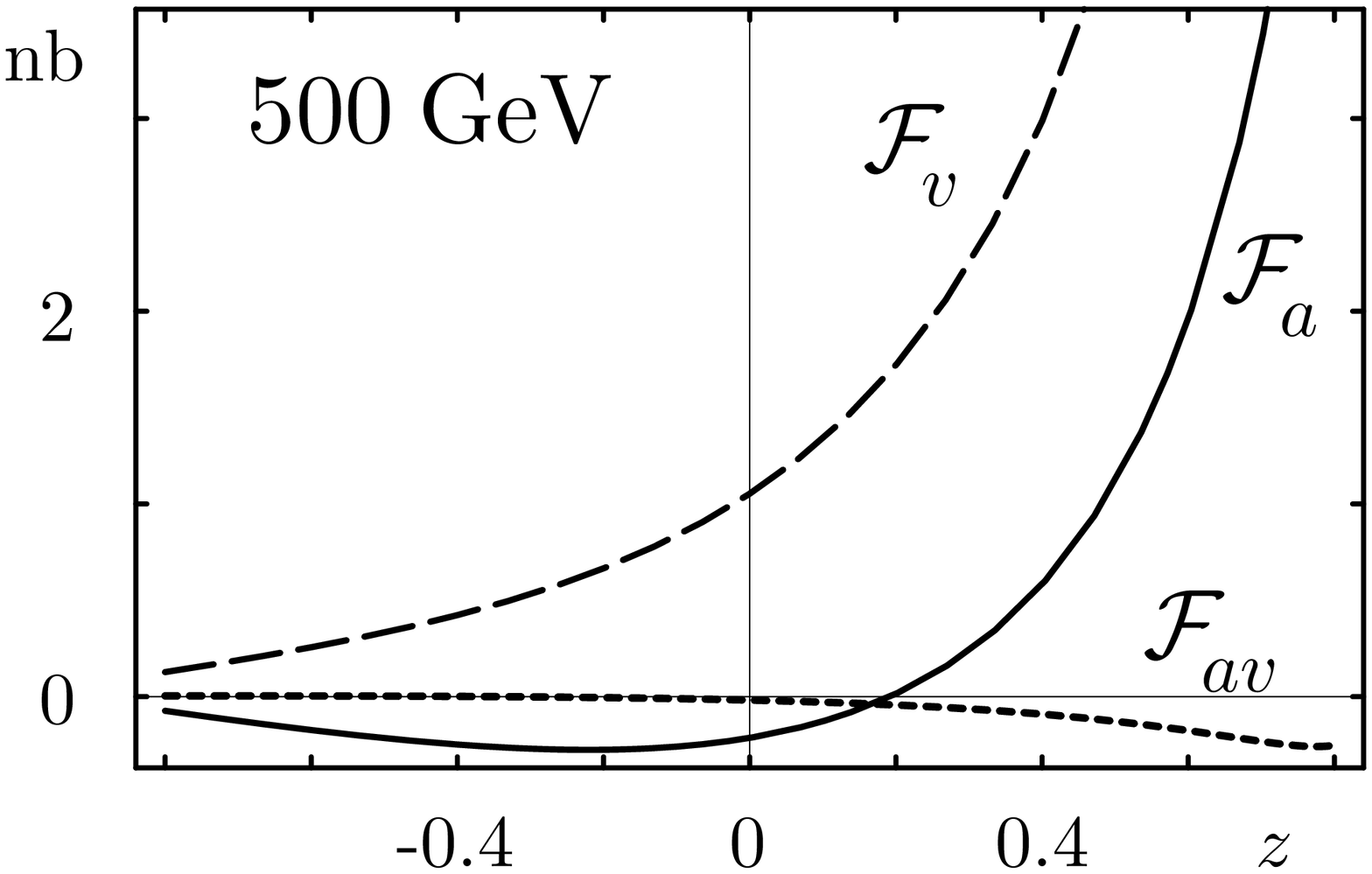}
\caption{Factors at $\bar{v}^2$, $\bar{a}^2$, and $\bar{a}\bar{v}$
in the differential cross-section $\Delta\,d\sigma/dz$
for $\sqrt{s}=200$ and 500 GeV.}
\end{figure}

The contributions to the cross-section of the process $e^+e^-\to
e^+e^- (\gamma)$ in the so-called improved Born approximation are
described by the tree-level plus one-loop diagrams with the
neutral vector boson exchange in the $s$ and $t$ channels. The
virtual states of the Abelian $Z'$ boson cause the differential
cross-section to differ from its SM value:
\begin{eqnarray}\label{3}
\Delta\,\frac{d\sigma}{dz} &=& \frac{d\sigma}{dz} -
\frac{d\sigma^{\rm SM}}{dz} =
 {\cal F}_{v}(\sqrt{s},z) \bar{v}^2 +
\nonumber\\&&+
{\cal F}_{a}(\sqrt{s},z) \bar{a}^2 + {\cal F}_{av}(\sqrt{s},z)
\bar{a}\bar{v}+\ldots,
\end{eqnarray}
where $z = \cos\theta$, $\theta$ -- scattering angle and we
introduced the dimensionless couplings
\begin{equation}\label{3a}
\bar{a} =
a_e\sqrt{\frac{\tilde{g}^2}{4\pi}\frac{m_Z^2}{m_{Z'}^2}},\quad
\bar{v} =
v_e\sqrt{\frac{\tilde{g}^2}{4\pi}\frac{m_Z^2}{m_{Z'}^2}}.
\end{equation}
Due to the correlations (\ref{2}) the cross-section contains only
three different combinations of the $Z'$ couplings $a_e^2$,
$v_e^2$, and $a_e v_e$. The factors ${\cal F}(\sqrt{s},z)$
standing at them depend on the SM couplings and particle masses.
For the purposes of the present investigation the values of
running couplings were taken at $\sqrt{s}\sim 200$ GeV. The dots
in Eq. (\ref{3}) denote the neglected  contributions of the box
diagrams and diagrams describing emission of soft photons in the
initial  and final states. These terms can be estimated to be of
the order of a few percents of the total and are inessential for
what follows. The explicit expressions for the tree-level
functions ${\cal F}(\sqrt{s},z)$ are adduced in the Appendix.
Their plots are shown in Fig. 1. As it is seen, each factor is
infinitely increasing at $z \to 1$. This is caused by the photon
exchange in the $t$-channel.

\begin{figure}
\centering
\includegraphics[bb= 0 0 500 350,width=70mm]{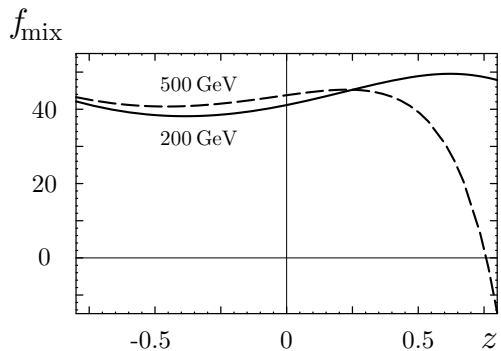}
\caption{The function $f_{\mathrm{mix}}(\sqrt{s},z)$ for the
center-of-mass energy $\sqrt{s}=200$ GeV (solid curve) and
$\sqrt{s}=500$ GeV (dashed curve).}
\end{figure}
\begin{figure}
\centering
\includegraphics[bb= 0 0 500 350,width=70mm]{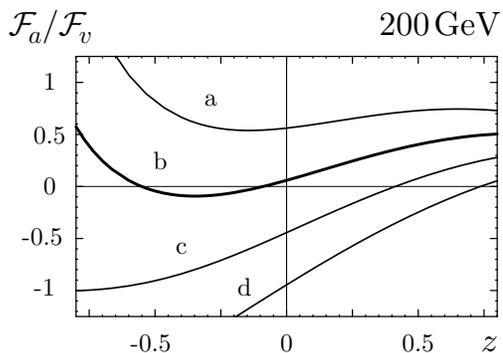}
\caption{The ratio ${\cal F}_{a}(z)/{\cal F}_{v}(z)$ of the
factors at $\bar{a}^2$ and $\bar{v}^2$ in the differential
cross-section $\Delta\,d\sigma/dz$ at various hypothetic values of
$\tilde{Y}_\phi/a_e$. The plots correspond to
$\tilde{Y}_\phi/a_e=-4$ (curve a), $-2$ (curve b, the Abelian
$Z'$), 0 (curve c), and 2 (curve d). The
center-of-mass energy is $\sqrt{s}=200$ GeV.}
\end{figure}

As it was mentioned above, the $Z$--$Z'$ mixing is expressed in
terms of the axial-vector coupling. So, it influences the factors
${\cal F}_a(\sqrt{s},z)$ and ${\cal F}_{av}(\sqrt{s},z)$. Without
taking into consideration the correlations (\ref{2}) the
contribution of the mixing to the cross-section has the form
\begin{equation}\label{5}
\left.\Delta\,\frac{d\sigma}{dz}\right|_{\mathrm{mix}} =
 -\frac{\alpha_{\mathrm{em}}}{144 s(1-z)}
 \frac{\tilde{g}^2 m_Z^2}{m_{Z'}^2}
 \tilde{Y}_{\phi}\,a_e\, f_{\mathrm{mix}}(\sqrt{s},z).
\end{equation}
The dimensionless function $f_{\mathrm{mix}}(\sqrt{s},z)$ is
adduced in the Appendix. For definiteness we show its angular
dependence at the center-of-mass energies 200 and 500 GeV in Fig.
2. The factor $f_{\mathrm{mix}}(\sqrt{s},z)$ is a non-zero
quantity which is almost an energy-independent constant at
$z<0.5$. The contribution of the mixing (\ref{5}) decreases when
the center-of-mass energy grows. At LEP2 energies the account of
the $Z$--$Z'$ mixing plays an important role. Actually, because of
the correlation (\ref{2}) the mixing terms essentially influence
the factors at axial-vector couplings in the cross-section. The
Abelian $Z'$ signal is charachterized by the value
$\tilde{Y}_\phi=-2a_e$. The function ${\cal F}_a(\sqrt{s},z)$
corresponding to this number is close to zero almost in all $z$
interval and qualitatively distinguishable among the factors for
other hypothetic values of $\tilde{Y}_\phi/a_e$ (see Fig. 3).
Hence, it is impossible to search for the signals of the Abelian
$Z'$ boson omitting the $Z'$--scalar coupling $\tilde{Y}_\phi$.
Setting $\tilde{Y}_\phi$ to zero, one obtains ${\cal
F}_a(\sqrt{s},z)$ which behavior does not reflect in general the
virtual state under consideration.

In order to study the deviations from the SM cross-section
inspired by the heavy virtual state of the Abelian $Z'$ boson, we
introduce the functions which are finite at all values of the
scattering angle. They are determined by means of dividing the
differential cross-section by some known monotonic function.

The factor ${\cal F}_{v}(z)$ is positive monotonic function of $z$
plotted in Fig. 4 for the center-of-mass energies $\sqrt{s}=200$
and 500 GeV. For intermediate energies the curves are in between
these very close located curves (in the chosen energy scale). This
type of behavior is typical for all other functions. So, in what
follows, as illustration, we shall present our results for these
two energies.
\begin{figure}
\centering
\includegraphics[bb= 0 0 380 270,width=70mm]{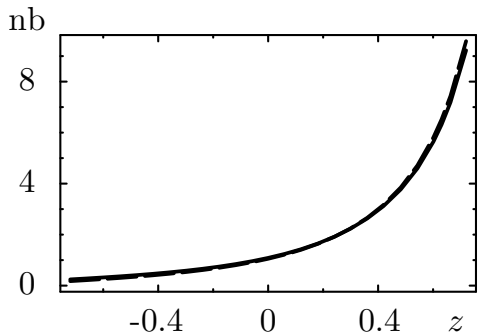}
\caption{Factor ${\cal F}_{v}(z)$ at the vector-vector
four-fermion coupling $\bar{v}^2$ in the deviation of the
differential cross-section of the Bhabha process from the SM
value, $\Delta\,d\sigma/dz$,
for $\sqrt{s}=200$ GeV (solid) and $500$ GeV (dashed).}
\end{figure}
Such a property allows one to choose ${\cal F}_{v}(z)$ as a
normalization factor for the differential cross section. Then the
normalized differential cross-section reads
\begin{eqnarray}\label{ncs}
\frac{d\tilde\sigma}{dz}&=&{\cal
F}_{v}^{-1}(\sqrt{s},z)\Delta\,\frac{d\sigma}{dz} =
\nonumber\\&&
 \bar{v}^2 +
F_{a}(\sqrt{s},z) \bar{a}^2 + F_{av}(\sqrt{s},z)\bar{a}\bar{v}
+\ldots ,
\end{eqnarray}
and the normalized factors are shown in Fig 5.
\begin{figure}
\centering
\includegraphics[bb= 0 0 500 350 ,width=70mm]{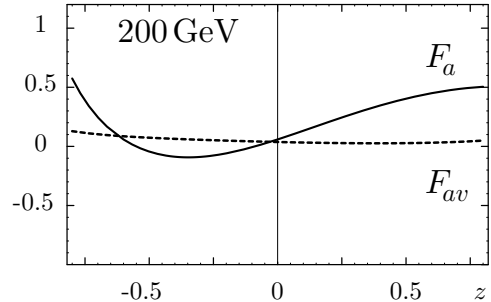}
\\
\includegraphics[bb= 0 0 500 350 ,width=70mm]{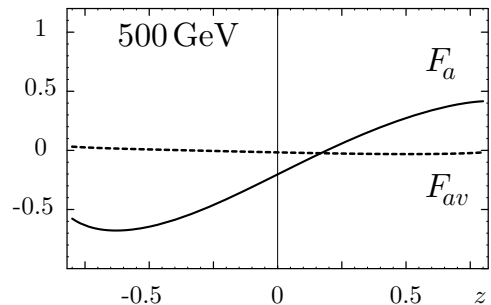}
\caption{Factors $F_{a}$ (solid) and $F_{av}$ (dashed) in the
normalized differential cross-section $\Delta\,d\tilde\sigma/dz$
for $\sqrt{s}=200$ and 500 GeV.}
\end{figure}

Now these factors are finite at $z\to 1$. Each of them in a
special way influences the differential cross-section.
\begin{enumerate}
\item
The factor at $\bar{v}^2$ is just the unity. Hence, the
four-fermion contact coupling between vector currents,
$\bar{v}^2$, determines the level of the deviation from the SM
value.
\item
The factor at $\bar{a}^2$ depends on the scattering angle in a
non-trivial way. It allows  to recognize the Abelian $Z'$ boson,
if the experimental accuracy is sufficient.
\item
The factor at $\bar{a}\bar{ v}$ results in small corrections.
\end{enumerate}

Thus, effectively, the obtained normalized differential
cross-section is a two-parametric function. In the next sections
we introduce the observables to fit separately each of these
parameters.

\section{Observables to pick out $\bar{v}^2$}

To recognize the signal of the Abelian $Z'$ boson by analyzing the
Bhabha process the differential cross-section deviation from the
SM predictions should be measured with a good accuracy. At
present, no such deviations have been detected at more than the
1$\sigma$ CL. In this situation it is resonable to introduce
integrated observables allowing to pick out $Z'$ signals by using
the most effective treating of available data. The observables
should be sensitive to the separate $Z'$ couplings. This admits of
search for the $Z'$ signals in different processes as well as to
perform global fits.

The deviation of the normalized differential cross-section
(\ref{ncs}) is (effectively) the function of two parameters,
$\bar{a}^2$ and $\bar{v}^2$. We are going to introduce the
integrated observables which determine  separately the
four-fermion couplings $\bar{a}^2$ and $\bar{v}^2$.

Let us first proceed with the observable for $\bar{v}^2$. After
normalization the factor at the vector-vector four-fermion
coupling becomes the unity. Whereas the factor at $\bar{a}^2$ is a
sign-varying function of the cosine of the scattering angle. As it
follows from Fig. 3, for the center-of-mass energy 200 GeV it is
small over the backward scattering angles. So, to measure the
value of $\bar{v}^2$ the normalized differential cross-section has
to be integrated over the backward angles. For the center-of-mass
energy 500 GeV the factor at $\bar{a}^2$ is already a
non-vanishing quantity for the backward scattering angles. The
curves corresponding to intermediate energies are distributed in
between two these curves. Since they are sign-varying ones at each
energy point some interval of $z$ can be chosen to make the
integral to be zero. Thus, to measure the $Z'$ coupling to the
electron vector current $\bar{v}^2$ we introduce the integrated
cross-section (\ref{ncs})
\begin{equation}\label{vobs}
\sigma_V = \int_{z_0}^{z_0+\Delta z} (d\tilde\sigma/dz)dz,
\end{equation}
where at each energy the most effective interval $[z_0,z_0+\Delta
z]$ is determined by the following requirements:
\begin{enumerate}
\item The relative contribution of the coupling $\bar{v}^2$ is
maximal. Equivalently, the contribution of the factor at
$\bar{a}^2$ is suppressed. \item The length $\Delta z$ of the
interval is maximal. This condition ensures that the largest
number of bins is taken into consideration.
\end{enumerate}

\begin{figure}
\centering
\includegraphics[bb= 91 3 322 234,width=65mm]{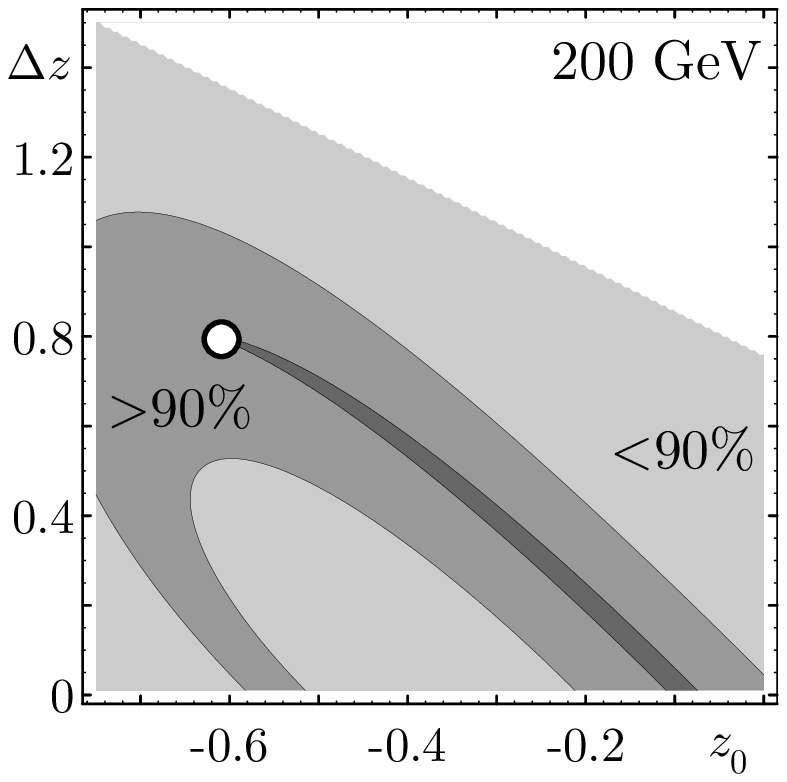}
\\
\includegraphics[bb= 91 3 322 234,width=65mm]{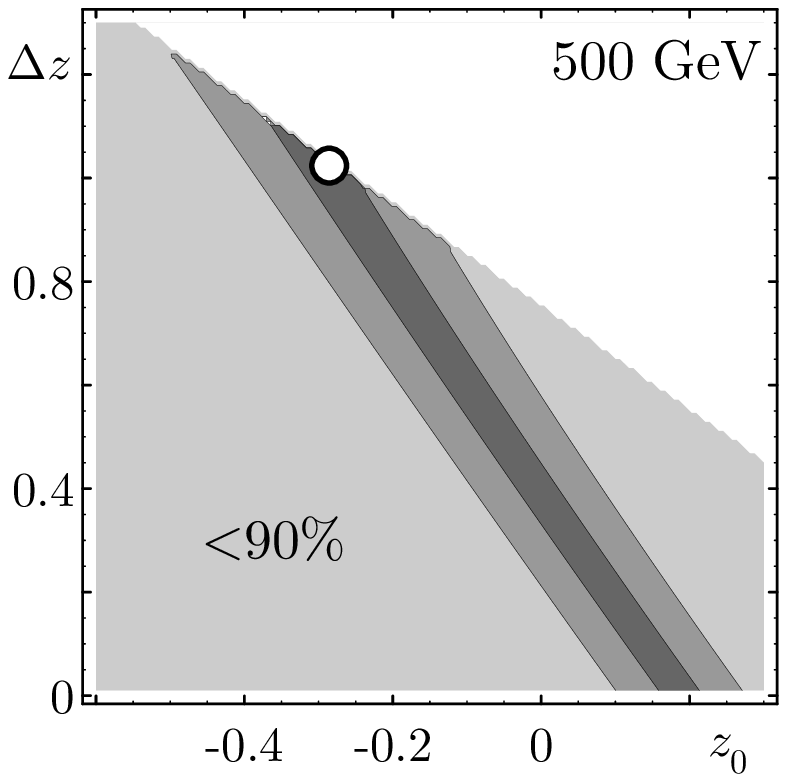}
\caption{Relative contribution of the factor at $\bar{v}^2$ as the
function of the left boundary of the angle interval, $z_0$, and
the interval length, $\Delta z$, at the center-of-mass energy 200
and 500 GeV. The shaded areas correspond to the contributions
$>95\%$
(dark), from 90\% to 95\% (midtone), and $<90\%$ (light).}
\end{figure}

The relative contribution of the factor at $\bar{v}^2$ is defined
as
\begin{equation}
\kappa_V= \frac{\Delta z}{\Delta z+\left|\int_{z_0}^{z_0+\Delta
z}F_a \,dz\right|+\left|\int_{z_0}^{z_0+\Delta
z}F_{av}\,dz\right|}
\end{equation}
and shown in Fig. 6 as the function of the left boundary of the
angle interval, $z_0$, and the interval length, $\Delta z$. In
each plot the dark area corresponds to the observables which
values are determined by the vector-vector coupling $\bar{v}^2$
with the accuracy $>95\%$. The area reflects the correlation of
the width of the integration interval $\Delta z$ with the choice
of the initial $z_0$ following from the mentioned requirements.
Within this area we choose the observable which includes the
largest number of bins (largest $\Delta z$). The corresponding
values of $z_0$ and $\Delta z$ are marked by the white dot on the
plots in Fig. 6. As the carried out analysis showed, the point
$z_0$ is shifted to the right with increase in energy whereas
$\Delta z$ remains approximately the same.

From the plots it follows that the most efficient intervals are
\begin{eqnarray}
&& -0.6<z<0.2,\quad\sqrt{s}=200\mbox{ GeV}, \nonumber\\&&
-0.3<z<0.7,\quad\sqrt{s}=500\mbox{ GeV}.
\end{eqnarray}
Therefore the observable (\ref{vobs}) allows to measure the $Z'$
coupling to the electron vector current $\bar{v}^2$ with the
efficiency $>95\%$.

\section{Observables to pick out $\bar{a}^2$}

In order to pick the axial-vector coupling $\bar{a}^2$ one needs
to eliminate the dominant contribution coming from $\bar{v}^2$.
Since the factor at $\bar{v}^2$ in the $d\tilde\sigma/dz$  equals
unity, this can be done by summing up equal number of bins with
positive and negative weights. In particular, the forward-backward
normalized differential cross-section appears to be sensitive
mainly to $\bar{a}^2$,
\begin{eqnarray}
\tilde\sigma_{\rm FB}&=&\int\nolimits_{0}^{z_{\rm
max}}dz\,\frac{d\tilde\sigma}{dz} -\int\nolimits_{-z_{\rm
max}}^{0}dz\,\frac{d\tilde\sigma}{dz} \nonumber\\ &=&
\tilde{F}_{a,\rm FB} \bar{a}^2 + \tilde{F}_{av,\rm
FB}\bar{a}\bar{v}.
\end{eqnarray}
The value $ z_{\rm max}$ is determined by the number of bins
included and, in fact, depends on the data set considered. The
LEP2 experiment accepted $e^+e^-$ events with $|z|<0.72$. In what
follows we take $z=0.7$ for definiteness.

The efficiency of the observable is determined as:
\begin{equation}
\kappa= \frac{|\tilde{F}_{a,\rm FB}|}{|\tilde{F}_{a,\rm
FB}|+|\tilde{F}_{av,\rm FB}|}.
\end{equation}
It can be estimated as $\kappa=0.9028$ for the center-of-mass
energy 200 GeV and $\kappa=0.9587$ for 500 GeV. To obtain this
value we  took typical $z_\mathrm{max}=0.7$. Thus, the observable
\begin{eqnarray}\label{aobs}
&& \tilde\sigma_{\rm FB}=0.224 \bar{a}^2 - 0.024
\bar{a}\bar{v},\quad\sqrt{s}=200\mbox{ GeV}, \nonumber\\&&
\tilde\sigma_{\rm FB}=0.472 \bar{a}^2 - 0.020
\bar{a}\bar{v},\quad\sqrt{s}=500\mbox{ GeV}
\end{eqnarray}
is mainly sensitive to the $Z'$ coupling to the axial-vector current $\bar{a}^2$.

Consider a usual situation when experiment is not able to
recognize the angular dependence of the differential cross-section
deviation from its SM value with the proper accuracy because of
loss of statistics. Nevertheless, a unique signal of the Abelian
$Z'$ boson can be determined. For this purpose the observables
$\int_{z_0}^{z_0+\Delta z}(d\tilde\sigma/dz)dz$ and
$\tilde\sigma_{\rm FB}$ must be measured. Actually, they are
derived from the normalized differential cross-section. If the
deviation from the SM is inspired by the Abelian $Z'$ boson both
the observables are to be positive quantities simultaneously. This
feature serves as the distinguishable signal of the Abelian $Z'$
virtual state in the Bhabha process for the LEP2 energies as well
as for the energies of future electron-positron colliders ($\geq
500$ GeV). The observables fix the unknown low energy vector and
axial-vector $Z'$ couplings to the electron current. Their values
have to be correlated with the bounds on $\bar{a}^2$ and
$\bar{v}^2$ derived by means of independent fits for other
scattering processes.

\section{LEP II data analysis}

Let us apply the introduced observables to analysis the
experimental data on differential cross-sections of the Bhabha
scattering. The LEP2 experiments have measured, in particular, the
differential cross-section of this process at a number of energy
points. Both the final and preliminary data are available in the
literature \cite{LEPdata}. These data provide an opportunity to
estimate possible Abelian $Z'$ virtual states.

In the present paper we take into consideration the differential
cross-sections measured by the L3 Collaboration at 183-189 GeV, by
the OPAL Collaboration at 189-207 GeV, as well as the
cross-sections obtained by the DELPHI Collaboration at energies
189-207 GeV.

First, we estimate the value of the vector coupling $\bar{v}^2$ by
using the observable (\ref{vobs}). At the LEP2 energies the
appropriate interval of the angular integration of the normalized
differential cross-section is $-0.6<z<0.2$. The results are shown
in Table \ref{tabl:fit} as `Fit 1'. The L3 and OPAL Collaborations
demonstrate positive values of $\bar{v}^2$ at the $1\sigma$ CL,
whereas the DELPHI Collaboration shows the $2\sigma$ positive
deviation. To constrain the value of the $Z'$ mass by the derived
bounds on the four-fermion coupling $\bar{v}^2$ some value of
$\tilde{g}$ should be fixed. Let us assume that the coupling
$\tilde{g}$ is of order of SM gauge couplings,
$\tilde{g}^2/(4\pi)\simeq 0.01-0.03$. Then the combined value of
$\bar{v}^2$ corresponds to $m_{Z'}=0.7-1.2$ TeV. It is interesting
to note that the mean values of $\bar{v}^2$ are a little larger
than those obtained for $\bar{a}^2$ for
$e^+e^-\to\mu^+\mu^-,\tau^+\tau^-$ processes \cite{YAF04}.
\begin{table}
  \centering
  \caption{Fits of the LEP2 data on the Bhabha process.
Fit 1 represents the value of $\bar{v}^2$ derived with the
observable $\sigma_V$ integrated in the angular interval
$z\in(-0.6, 0.2)$. In Fit 2 the axial-vector coupling $\bar{a}^2$
is fitted by the analysis of the observable
$\tilde{\sigma}_\mathrm{FB}$. In Fit 3 the value of
$\bar{a}^2=3.6\times 10^{-5}$ from the analysis of the
$e^+e^-\to\mu^+\mu^-$ scattering process is taken, and the value
of vector coupling $\bar{v}^2$ is found by fitting the total
cross-section ($|z|<0.72$).}\label{tabl:fit}
\begin{tabular}{|c|r|r|r|} \hline
 $\sqrt{s}$, GeV & $\bar{v}^2$, Fit 1 & $\bar{a}^2$, Fit 2 & $\bar{v}^2$, Fit 3  \\
 \cline{2-4} & \multicolumn{3}{|c|}{$\times 10^{-4}$} \\ \hline
 \multicolumn{4}{|c|}{DELPHI}\\ \hline
 189 &  $1.0\pm 2.8$ & $-23\pm 16$ &  $0.9\pm 2.0$ \\
 192 &  $3.1\pm 6.5$ &   $4\pm 34$ &  $2.5\pm 4.4$ \\
 196 &  $5.1\pm 3.9$ &   $1\pm 20$ &  $5.0\pm 2.6$ \\
 200 &  $6.1\pm 3.7$ & $-33\pm 20$ &  $3.0\pm 2.4$ \\
 202 & $-1.6\pm 4.6$ &   $6\pm 25$ &  $0.8\pm 3.2$ \\
 205 &  $2.7\pm 3.4$ & $-27\pm 19$ &  $0.8\pm 2.3$ \\
 207 &  $1.4\pm 2.9$ &  $-1\pm 15$ & $-1.6\pm 1.9$ \\
 Combined & $2.39\pm 1.35$ & $-12\pm 7$ & $1.2\pm 0.9$ \\
\hline\multicolumn{4}{|c|}{OPAL}\\ \hline
 189 & $-2.8\pm 2.2$ &   $1\pm 12$ &  $0.2\pm 1.5$ \\
 192 & $-2.4\pm 6.4$ &  $20\pm 34$ &  $1.6\pm 4.2$ \\
 196 &  $0.3\pm 3.1$ &   $1\pm 15$ & $-0.1\pm 2.0$ \\
 200 &  $3.4\pm 3.1$ &  $-1\pm 16$ &  $2.9\pm 2.1$ \\
 202 &  $9.1\pm 5.6$ &  $-9\pm 27$ &  $3.5\pm 3.4$ \\
 205 &  $7.2\pm 3.1$ & $-11\pm 16$ &  $3.9\pm 2.1$ \\
 207 &  $1.5\pm 2.3$ &  $-4\pm 13$ &  $1.7\pm 1.6$ \\
 Combined & $1.33\pm 1.14$ & $-2\pm 6$ & $1.6\pm 0.8$ \\
 \hline\multicolumn{4}{|c|}{L3}\\ \hline
 183 &  $3.7\pm 5.8$ & $-36\pm 35$ &  $3.8\pm 5.0$ \\
 189 &  $1.9\pm 3.2$ &   $7\pm 18$ & $-1.6\pm 2.6$ \\
 Combined & $2.36\pm 2.82$ & $-2\pm 16$ & $-0.4\pm 2.3$ \\
 \hline\multicolumn{4}{|c|}{COMBINED}\\ \hline
     &  $1.82\pm 0.83$ & $-6\pm 4$ & $1.3\pm 0.6$ \\
 \hline
\end{tabular}
\end{table}

The second fit (`Fit 2') estimates the observable (\ref{aobs})
related to the value of $\bar{a}^2$. Since in the Bhabha process
the effects of the axial-vector coupling are suppressed with
respect to those of the vector coupling, we expect much larger
experimental  uncertainties for $\bar{a}^2$. Indeed, the LEP2 data
lead to the significant errors for  $\bar{a}^2$ of order
$10^{-3}-10^{-4}$ (See Table \ref{tabl:fit}). The mean values are
negative numbers which are too large to be interpreted as a
manifestation of some heavy virtual state beyond the energy scale
of the SM.

Finally, we performed also the fit of $\bar{v}^2$, assuming the
value of the $Z'$-induced axial-vector coupling from
$e^+e^-\to\mu^+\mu^-$ process \cite{YAF04}. In this case, putting
$\bar{a}^2=0.000037$ and computing the total cross-section
($|z|<0.72$), we obtain the observable which depends on one
parameter $\bar{v}^2$, only. This parameter can be easily fitted.
As it follows from Table \ref{tabl:fit}, the results are close to
those based on the observable (\ref{vobs}) for the vector
coupling.

Thus, the LEP2 data constrain the value of $\bar{v}^2$ at the
$2\sigma$ CL which could correspond to the Abelian $Z'$ boson with
the mass of the order 1 TeV. In contrast, the value of $\bar{a}^2$
is a large negative number with a significant experimental
uncertainty. This can not be interpreted as a manifestation of
some heavy virtual state beyond the energy scale of the SM.

\section{Conclusion}

In the present paper we have introduced new observables for
model-independent searches for the heavy Abelian $Z'$ boson in the
Bhabha process at the energies of the LEP and future linear
electron-positron colliders. They are based on the specific
correlations (\ref{2}) existing at low energies between the vector
and the axial-vector $Z'$ couplings to fermions and therefore
uniquely identify this virtual state. If, for instance, one does
not take into consideration these relations, no estimation of the
$v^2_f$ could be derived at all.

It is interesting to compare the results on the $Z'$ search in the
Bhabha processes with that of  the processes $e^+ e^-\to
\mu^+\mu^-,\tau^+\tau^-$ investigated already in Ref.
\cite{YAF04}. Cross-sections for those processes involve quadratic
combinations of electron couplings $a_e$, $v_e$ and muon (tau)
couplings $a_\mu$, $v_\mu$ ($a_\tau$, $v_\tau$). Due to Eq.
(\ref{2}) the axial-vector coupling is universal,
$a_e=a_\mu=a_\tau=a_l$. Therefore, there is the sign-definite term
$a^2_l$ in the cross-section. A more simple kinematics of the
processes $e^+ e^-\to \mu^+\mu^-,\tau^+\tau^-$ allows to introduce
a one-parametric observable which is directly related to the
coupling $a_l^2$. Just the sign of the observable serves as the
signal of the Abelian $Z'$ in this case. The following estimate
has been derived:
\begin{eqnarray} \label{2003}
\mu^+\mu^-:&&\bar{a}^2_l=(3.66\pm 4.89) \times 10^{-5},
\nonumber\\ \tau^+\tau^-:&& \bar{a}^2_l=(-2.66\pm 6.43) \times
10^{-5}.
\end{eqnarray}
The dimensionless coupling $\bar{a}_l$ is related to $a_l$ by
means of Eq. (\ref{3a}) and actually identical to the parameter
$\bar{a}$ admitted in the present paper. The more precise
$\mu^+\mu^-$ data demonstrate the $1\sigma$ deviation whereas the
$\tau^+\tau^-$ data do not specify the sign of the observable. As
the value of $v^2_f$ is concerned, it remained completely
unrestricted in that case because of the kinematics specifics.
Thus, the results derived in the present paper from the analysis
of the Bhabha process are complementary. We see also that the
accuracy of the result is of $2\sigma$ CL vs. $1\sigma$ for $e^+
e^- \rightarrow l^+ l^-$.

There is a simple relation of the parameters $\bar{a}$, $\bar{v}$
used in the present paper to the four-fermion couplings
$\epsilon_{AA}$, $\epsilon_{AV}$, and $\epsilon_{VV}$ admitted in
the reports of the Electroweak Working Group \cite{EWWG}. They are
just normalized by different factors and related as
$\epsilon_{AA}=-\bar{a}^2 m^{-2}_Z/4$,
$\epsilon_{AV}=-\bar{a}\bar{v} m^{-2}_Z/4$, and
$\epsilon_{VV}=-\bar{v}^2 m^{-2}_Z/4$. The description of the
Abelian $Z'$ virtual state in the Bhabha process requires both
$\epsilon_{VV}$ and $\epsilon_{AA}$. None of these couplings
should be set to zero in searching for $Z'$ signals. The separate
constraints on $\epsilon_{VV}$ or $\epsilon_{AA}$ could be
performed by means of observables $\sigma_V$ and
$\tilde\sigma_{\rm FB}$.

Let us confront our results with that in Ref. \cite{bourilkov}
where the `helicity model fit' was applied for the total cross
sections of the Bhabha process. The most interesting result of
that analysis is the observation that for the AA-model the
$2\sigma$ deviation from the SM has been derived. For other
`models' no distinguishable deviations were observed. Since our
analysis is based on the properties of differential cross
sections, no direct comparisons can be done. We just mention that
in the paper \cite{YAF04} we showed that the AA-helicity model is
mainly responsible for the signal of the Abelian $Z'$ gauge boson.
So, one may speculate that this fact found its implementation in
the significant $2\sigma$ deviations.

As a general conclusion we would like to stress that the
observables to search for the Abelian $Z'$ signal in Bhabha
process can be introduced at the LEP2 energies ($\sqrt{s}\simeq
200$ GeV) as well as at the energies of future electron-positron
colliders ($\sqrt{s}\geq 500$ GeV). They are determined by
normalized differential cross-section and measure two
sign-definite $Z'$ couplings, $v^2$ and $a^2$. The positive signs
of the observables are the characteristic feature of the Abelian
$Z'$ virtual state. The corresponding values of $Z'$ couplings
could be compared with those obtained in other independent
scattering processes.

In the Bhabha process the Abelian $Z'$ effects are dominated by
the vector coupling $v^2$. This process provides mainly
constraints on the $Z'$-induced vector-vector four-fermion
coupling. Therefore it gives complementary information to the
constraints on the axial four-fermion coupling $a^2$ based on the
analysis of $e^+e^-\to\mu^+\mu^-,\tau^+\tau^-$. From the carried
out investigations we conclude that the $Z'$ boson is expected to
have the mass of the order $1 - 1.2$ TeV and has a good chance to
be discovered at colliders of next generation.

\section*{Acknowledgement}

The authors are grateful to A. Babich and A. Pankov for numerous
discussions. AG thanks ICTP (Trieste, Italy) for a hospitality at
the High Energy Section when the final version of the paper was
prepared. This work was partially supported by the grant No
F7/296-2001 from the Foundation for Fundamental Researches of the
Ministry of Education and Science of Ukraine.

\section*{Appendix}

Below we adduce explicitly the factors ${\cal F}_{v}$, ${\cal
F}_{a}$, ${\cal F}_{av}$ at the $Z'$-induced couplings
$\bar{v}^2$, $\bar{a}^2$ and $\bar{a}\bar{v}$ in the differential
cross-section $\Delta\,d\sigma/dz$ (Eq. (\ref{3})). These factors
are plotted in Fig. 1 at the so-called imroved-Born level. The
complete expressions for ${\cal F}_{v}$, ${\cal F}_{a}$, and
${\cal F}_{av}$ are quite bulky. They contain the terms of
different order in the small parameters $\mu_z=m^2_Z/s$ and
$1-4\sin^2\theta_W\simeq 0.08$ ($\theta_W$ is the Weinberg angle).
The terms of order $1-4\sin^2\theta_W$ influence the total values
by a few percents. Therefore we present the contributions of
$\mu_z=m^2_Z/s$ only:
\begin{eqnarray}
&& {\cal F}_{v}(\sqrt{s},z)=
 \frac{\pi\alpha_{\mathrm{em}}[1+O(1-4\sin^2\theta_W,m^2_e/s)]}
 {24m_Z^2(1-z)(1-\mu_z)(1-z+2\mu_z)^2}
 \times\nonumber\\ &&\times
\big[
 22 + 73 \mu_z + 4 \mu_z^2 - 108 \mu_z^3
 -z (14 - 19 \mu_z - 64 \mu_z^2
    +\nonumber\\ &&\quad
 + 108 \mu_z^3)
 -2 z^2 (10 + 3 \mu_z - 68 \mu_z^2 + 18 \mu_z^3)
 +\nonumber\\ &&\quad
 +2 z^3 (2 - 21 \mu_z + 24 \mu_z^2 - 18 \mu_z^3)
 -\nonumber\\&&\quad
 -z^4 (2 + 35 \mu_z - 36 \mu_z^2)
 +z^5 (10 - 9 \mu_z)
%
%
\big],
 \nonumber\\
&& {\cal F}_{a}(\sqrt{s},z)=
 \frac{\pi\alpha_{\mathrm{em}}[1+O(1-4\sin^2\theta_W,m^2_e/s)]}
 {24m_Z^2(1-z)(1-\mu_z)^2(1-z+2\mu_z)^3}
 \times\nonumber\\ &&\times
\big[
 -3 (2 - 3 \mu_z - 53 \mu_z^2 - 2 \mu_z^3 + 132 \mu_z^4 - 40 \mu_z^5)
 +\nonumber\\ &&\quad
 +6 z (6 - 3 \mu_z - 36 \mu_z^2 + 61 \mu_z^3 + 20 \mu_z^4 - 12 \mu_z^5)
 -\nonumber\\ &&\quad
 -z^2 (42 - 131 \mu_z - 135 \mu_z^2 + 196 \mu_z^3 + 64 \mu_z^4 + 216 \mu_z^5)
 \nonumber\\ &&\quad
 -4 z^3 (6 + 51 \mu_z - 2 \mu_z^2 - 43 \mu_z^3 - 82 \mu_z^4 + 6 \mu_z^5)
 +\nonumber\\ &&\quad
 +z^4 (54 + 47 \mu_z - 99 \mu_z^2 - 162 \mu_z^3 +12 \mu_z^4)
 -\nonumber\\ &&\quad
 -2 z^5 (6 - 15 \mu_z - 8 \mu_z^2 - 3 \mu_z^3)
 -z^6 (6 - 5 \mu_z + 3 \mu_z^2)
\big],
 \nonumber\\
&& {\cal F}_{av}(\sqrt{s},z)=
 \frac{\pi\alpha_{\mathrm{em}}O(1-4\sin^2\theta_W,m^2_e/s)}
 {6m_Z^2(1-z)(1-\mu_z)^2(1-z+2\mu_z)^3}.
%
%
 \nonumber
\end{eqnarray}

We also adduce the leading-order expression for the function
$f_{\mathrm{mix}}(\sqrt{s},z)$ entering Eq. (\ref{5}):
\begin{eqnarray}
&& f_{\mathrm{mix}}(\sqrt{s},z) =
\frac{1}{(1-z+2\mu_z)^2(1-\mu_z)^2}
 \times\nonumber\\ &&\times \big[
 45 + 81 \mu_z - 243 \mu_z^2 + 90 \mu_z^3
 -9z (9 - 12 \mu_z
 -\nonumber\\ &&\quad
  -8 \mu_z^2 +6 \mu_z^3)
 +z^2 (18 - 18 \mu_z + 114 \mu_z^2 - 162 \mu_z^3)
 -\nonumber\\ &&\quad
 -z^3 (18 + 108 \mu_z - 192 \mu_z^2 +18 \mu_z^3)
 +z^4 (33 - 63 \mu_z
 +\nonumber\\ &&\quad
  +9 \mu_z^2)
 +3z^5
 \big]
\big[1+O(1-4\sin^2\theta_W,m^2_e/s)\big]\nonumber.
\end{eqnarray}
This quantity is finite for the all values of scattering angle.

\end{document}